\documentclass[11pt,a4paper]{article}
\usepackage[margin=1in]{geometry}
\usepackage[T1]{fontenc}
\usepackage{lmodern}
\usepackage{microtype}
\usepackage{fancyvrb}

\usepackage{parskip}
\usepackage{datetime}
\usepackage{amsmath,amssymb,amsthm}
\usepackage{stmaryrd}
\usepackage{mathtools}
\usepackage{enumitem}
\usepackage{hyperref}
\usepackage[nameinlink,capitalise]{cleveref}
\usepackage{xcolor}
\usepackage{tikz}
\usepackage{tikz-cd}
\usetikzlibrary{arrows.meta,automata,calc,positioning,shapes.geometric}

\usepackage[authoryear,round]{natbib}
  
\usepackage{graphicx}

 \usepackage{mathptmx}      

\tikzcdset{
  scale cd/.style={
    every label/.append style={scale=#1},
    cells={nodes={scale=#1}}
  }
}
\tikzset{
  box/.style={draw, rounded corners, align=center, fill=blue!10, inner sep=6pt},
  boxy/.style={draw, rounded corners, align=center, fill=yellow!15, inner sep=6pt},
  arrow/.style={-Latex, line width=0.9pt}
}

 \title{AI as a component in the action research tradition of learning-by-doing}      
\author{Ian Benson and Alexei Semenov}
\date{\today}

\begin{document}
\maketitle

\begin{abstract}

We consider learning mathematics through action research, hacking, discovery, inquiry, learning-by-doing as opposed to the ‘instruct and perform’, ‘industrial’ model of the 19th century. A learning model based on self-awareness, types, functions, structured drawing and formal diagrams addresses the weaknesses of drill and practice and the pitfalls of statistical prediction with Large Language Models. 

In other words, we build mathematics/informatics education on the activity of a professional mathematician in mathematical modelling and designing programs. This tradition emphasises the role of dialogue and doing mathematics. In the Language/Action approach the teacher designs `mathematising situations' that scaffold previously encountered, or ‘not-known-how-to-solve’ problems for the learner while teachers and teacher/interlocutors supervise the process.

A critical feature is the written-oral dialogue between the learner and the teacher. As a rule, this is 1-1 communication. The role of the teacher-interlocutor, a ‘more knowledgeable other', is mostly performed by a more senior student, 1 per 5–7 pupils. After Doug Engelbart we propose the metaphor of human intellect augmented by digital technologies such as interactive development environments or  AI. Every human has their bio and digital parts. The bio part of the learner reacts to  their work through dialogue in the mind. The digital part poses questions, interprets code and proposes not necessarily sound ideas.

\end{abstract}


\textbf{Who did what} IB and AS devised the title, abstract and manuscript structure. AS contributed the role of philosophers and educationists on learning through dialogue and learning by doing and the history of mathematics circles. IB reported on mathematics education in England and the replication and extension of the Cuisenaire-Gattegno approach. IB applied Language/Action theory to the specification of a new technical infrastructure that integrates Interactive Development Environments and generative AI agents into the mathematics learning network. Both authors reviewed the manuscript.

\textbf{Citation}
This paper was developed from a presentation on \textit{AI as a component of an action research model of learning} at the 2024 Cambridge Generative AI in Education Conference, October 2024. It was submitted for publication in a Special Issue of the Cambridge Journal of Education.

\pagebreak

\section{Introduction}

The introduction of generative AI systems in mathematics teaching requires well understood and functioning educational processes. At the same time teaching needs to move beyond cycles of `instruct--perform.'  In Section~\ref{fts} we examine the state of the mathematical education system in England, with particular attention to the processes establishing curriculum content, pedagogy and performance management. We identify a deficit in mathematical thinking.

In Section~\ref{beyond} we offer a resume of the theoretical and philosophical alternatives to `teacher-led' instruction and `mathematics mastery.'  Although reform is represented here by non-traditional text book series, \textit{Mathematics and Informatics: Vols 1-4} \citep{Soprunova:2021aa} and \textit{Gattegno Mathematics: Vols 1-7 (formerly Mathematics with Numbers in Colour)} \citep{Gattegno:1963cr}, the primary distinction between reformed and traditional mathematics teaching practice is in our emphasis on action research. Action research is a classroom based, iterative process where researchers and practitioners collaborate in lesson design,  process validation and peer-review. Where the industrial division of labour in mass education demands a strict separation between curriculum and resource developers and teachers the action research tradition emphasises teacher agency and the harnessing of the learner's inherent mental processes.

In Section~\ref{action} we describe the action research initiative of the Belgian teacher Georges Cuisenaire and his collaborator Caleb Gattegno. Cuisenaire's invention, colour coded rods, were influential as manipulative aids in learning arithmetic and were popular throughout the Anglophone and Francophone world. They prompted Gattegno and his colleagues to establish, in 1952, the UK Association for Teaching Aids in Mathematics (ATAM), later to become the Association of Teachers of Mathematics (ATM)  now a component of the Association for Mathematics in Education. They developed a curriculum and text books that based mathematics learning on an enriched technical vocabulary: including structured drawing (to create visual objects), sets, functions and morphisms. Section~\ref{haskell} adds a computational dimension to this endeavour by including software enabled type inference. This directs the learner's attention to the multiple infinitesimal steps that comprise mathematical thought. Section~\ref{others} describes related work incorporating computing concepts such as string, set and bag into teaching materials and visual programming environments such as Karel the Robot and Monty Karel. 

Section~\ref{codesign} we propose a reconfiguration of the mathematics classroom as a learning network. Section~\ref{conclude} concludes the paper with reflections on how the transition to this new way of working might be co-designed with teachers in a gradual evolution to a new model.

ChatGPT and Claude were used in the summary description of the primary school curriculum in Section 2 and in the overview of the pedagogies for mathematics education in Section 3.

%

\section{Mathematics Education in England}
\label{fts}

Educational process design in England is a mix of top-down managerial thinking with teacher agency increasingly constrained by management structures such as Multi-Academy Trusts.  The Department for Education (DfE) administers standardised tests, school performance league tables and other accountability measures. In service teacher training and lesson planning is hierarchical, with training often `cascading' from `hubs.'  A national curriculum regulates the content of teaching, national assessments and public examinations. The \textit{counting-led} primary school curriculum is organised as two stages:

\textbf{KS1 (Years 1–2)}: Aims to build `number sense' and fluency. Counting and place value (to 100+), secure addition/subtraction with written and mental methods, early multiplication/division via equal groups/arrays, simple fractions (½, ¼), measures (length, mass, time, money), properties of 2D/3D shapes, and reading simple charts/tables -- with an emphasis on reasoning and solving everyday problems.

\textbf{KS2 (Years 3–6)}: Aims to extend and secure the four operations (including long multiplication/division), larger integers, negative numbers, powers of 10 and rounding. Fractions, decimals, percentages and ratio are connected and used to solve multi-step problems. \textit{Geometry deepens (angles, polygons, circles parts, symmetry, coordinates, perimeter/area/volume), measures (metric/imperial), statistics (bar/line graphs, tables) and introductory algebraic thinking (simple formulae, unknowns)}.

Students may be graded in national assessments as \textit{working at the expected level} at the end of Year 6 without attempting any questions of the italised material encountered in years 5 and 6  \citep{Land:2024aa,Keating:2026aa}.

A recent study by \cite{Marks:2023aa}  reports that as a result of misgivings around textbook-schemes -- or even other mathematics schemes -- many primary
teachers opt to curate their mathematics curriculum resources, developing a ‘patchwork’ through creating their own or sourcing resources from various places. Survey data suggests that average primary school teachers spend between 1 and 3 hours a week searching online for supplementary mathematics curriculum resources. Concern has previously been raised around the patchwork of curriculum resources that teachers acquire. They may be of inherently dubious quality, with a particular concern about the poor quality and limited cognitive demand of resources uploaded to repositories for teachers to share curriculum resources.  

In the past teachers have had a high degree of agency in how any curriculum resource is used. This freedom is increasingly circumscribed as schools amalgamate into Multi-Academy Trusts who often have preferred suppliers of commercial curriculum mapping resources such as White Rose. The DfE is now promoting its own `AI-powered' lesson design assistant through its free digital curriculum resource portal -- Oak National Academy. Schools increasingly set and mark assignments with gamified computer based instruct-perform workbooks, with school based student league tables. In due course national assessments and examinations themselves are expected to move online.

Readers will find a comprehensive account of `evidence-based' justifications for this status quo in \cite{Gibb:2025aa}. In practice the DfE and other agencies  involved in curriculum planning, such as the Royal Society Mathematics Futures Board, have limited ambitions for learners, teachers and schools \citep{Group:2023aa}. In 2025 26\% of primary pupils failed even to achieve `expected level.' Similarly, despite being the only school subject that has two tiers of GCSE examinations, it was possible in 2025 with the AQA and Pearson examining boards to be awarded a standard mathematics pass (level 4) by answering only  22--25\% of the Higher Tier questions correctly, or 60\% of the Foundation Tier. The 26\% of students who fail to reach expected standards at the end of primary school increases to 42\% who fail to reach a standard pass at GCSE.

\section{Mathematical Thinking is Needed}
\label{beyond}
%

It is against this background of low expectations, and failed educational performance, that we offer a resume of the theoretical and philosophical alternatives to '`teacher-led instruction' and `mathematical mastery.'  There is a long tradition of  educationists who have emphasised the value of learning mathematics by doing and by oral and written dialogue. Mathematicians have themselves reflected on the best way to educate practitioners, and more recently they have drawn on concepts from computer science to enrich learning and teaching. 

The question of how mathematical knowledge is best learned and transmitted has occupied educators, philosophers, and mathematicians for centuries. From early modern pedagogues such as John Amos Comenius to twentieth-century figures including R.\,L.\ Moore, George Pólya, and Paul Halmos. A consistent thread runs through their reflections: mathematics cannot be fully assimilated through passive listening or rote memorisation. Instead, it requires active engagement with problems, proofs, and arguments. 

A parallel theme arises when one turns from the activity of doing mathematics to the modes of communication that make mathematics a public, shareable discipline. Thinkers such as M.\,M.\ Bakhtin, Caleb Gattegno, Vladimir Bibler, and Rupert Wegerif have insisted that oral and written dialogues are not secondary, but constitutive of what we call ``mathematical reality.'' In their different ways, they demonstrate that the act of speaking, writing, and negotiating meaning is inseparable from the educational process.

Mathematicians and cognitive scientists including Jacques Hadamard, V. V. Davydov, George Lakoff, Charles Wells and Caleb Gattegno bring a scientific perspective to the design of instructional strategies. Hadamard highlighted the mental activity that precedes mathematical expression. Davydov indentified this phase as \textit{abstract thinking}, understood as an awareness of the relations that constitute a concept. Lakoff posed a contrast between \textit{categories} seen as abstract, shared, properties and \textit{categorising} that relies on human determined basic-level concepts and social interactions. Wells brings these concerns up to date when he argues that computing concepts and practices have much to offer mathematical education. 

Recent developments in K-12 computer science, generative AI and in mathematics circles in Britain suggests that educational mathematics software supported mathematics circles can be a most important model for the mathematics education system in mass schooling for all students.

\subsection{Learning by Doing: Comenius, Moore, Pólya, Halmos}
Comenius’s \emph{Great Didactic} emphasized a pedagogy that moves learners from sensory and practical experiences toward abstraction. He conceived of the school as a workshop, in which pupils do not merely observe but instead learn to speak by speaking and to reason by reasoning \citep{Comenius1907}. In mathematics education, this implies that early engagement with tangible objects, patterns, and everyday problems lay the foundation for formal concepts. 

R.\,L.\ Moore advanced this principle within the context of advanced mathematics. Through his eponymous method, Moore gave students definitions and theorems but withheld proofs, compelling them to generate arguments independently \citep{MahavierMore}. This approach generated profound independence and creativity, but it also raised concerns about inclusivity. Nevertheless, the Moore method remains historically influential, and contemporary inquiry-based learning can be seen as a modified continuation of Moore’s attempt to make students genuine participants in mathematical research.

George Pólya, in contrast, elaborated a more general theory of problem solving. In \emph{How to Solve It}, he proposed a four-step process: understanding the problem, devising a plan, carrying it out, and looking back \citep{Polya1945}. Central to his approach were \emph{heuristics}, or general strategies such as drawing figures, considering simpler cases, or working backwards. Pólya argued that these methods are best acquired by practice; that is, by working on real problems rather than simply learning procedures

Paul Halmos echoed these sentiments from the vantage point of a professional mathematician. His well-known assertion that ``the best way to learn is to do; the worst way to teach is to talk'' encapsulates his conviction that mathematics education should be organised around active engagement \citep{HalmosMoore}. He advocated for courses built on student presentations, problem solving, and critical discussion, arguing that mathematical understanding crystallises only through active practice.

\subsection{Learning through Dialogue: Bakhtin, Bibler, Wegerif}

Bakhtin’s dialogic philosophy underscores the idea that meaning is not monologic but emerges from the interaction of multiple voices. His distinction between authoritative discourse and internally persuasive discourse illustrates how learners appropriate mathematics: not by passive reception, but by re-voicing concepts in their own words \citep{BakhtinDialogic,BakhtinPoetics}. Oral dialogue allows the spontaneous negotiation of meaning, while written genres such as proofs and definitions lend permanence. This insight suggests that teachers should create conditions where students’ voices are genuinely heard, allowing them to negotiate between the authoritative voice of mathematics and their own developing understandings.

Vladimir Bibler extended the notion of dialogue into the philosophy of culture. His ``school of the dialogue of cultures'' presented knowledge as something forged at the boundaries of cultural traditions and intellectual frameworks \citep{Bibler:2009aa}. In mathematics, this insight highlights that concepts and methods emerge not in isolation but in conversation with history, culture, and competing worldviews. 

Rupert Wegerif brings Bakhtin and Bibler’s philosophical insights into  contemporary educational thinking. He characterises learning as the expansion of ``dialogic space,'' where students engage not only with peers and teachers but also with cultural voices and the ``Infinite Other'' \citep{wegerif2013}. In practice, this means cultivating classrooms where both talk and writing are vehicles for opening perspectives, questioning assumptions, and connecting local problem solving with a broader cultural narrative. Wegerif’s dialogic theory suggests that education in mathematics should not only teach correct answers but also nurture the capacity to inhabit multiple perspectives and to engage responsibly with the wider cultural significance of mathematical thinking.

\subsection{Instructional Strategy and Tactics: Hadamard, Davydov, Lakoff, Wells}

Jacques Hadamard’s \emph{The Psychology of Invention in the Mathematical Field} contributes a complementary perspective by examining the phenomenology of mathematical creativity. Based on testimonies from leading mathematicians, Hadamard concluded that invention often proceeds through nonverbal imagery, sensations of movement, and other pre-verbal forms of thought \citep{Hadamard1945}. Verbal expression, whether oral or written, typically enters later in the process, serving to crystallise, verify and communicate.

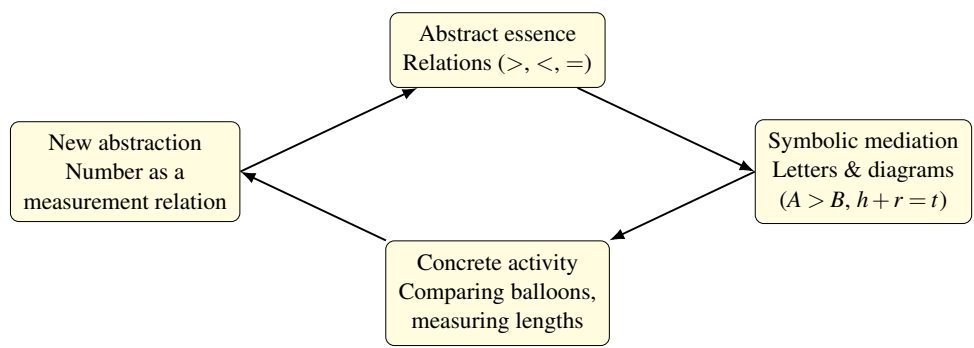
\begin{figure}[h!]
\centering
\resizebox{0.8\textwidth}{!}{%
\begin{tikzpicture}
  \coordinate (A) at (0, 2);
  \coordinate (S) at (4.2, 0);
  \coordinate (C) at (0,-2);
  \coordinate (N) at (-4.2,0);

  \node[boxy] (Abstract) at (A)
    {Abstract essence\\Relations ($>$, $<$, $=$)};
  \node[boxy, right=0pt of S] (Symbolic)
    {Symbolic mediation\\Letters \& diagrams\\($A>B$, $h+r=t$)};
  \node[boxy] (Concrete) at (C)
    {Concrete activity\\Comparing balloons,\\measuring lengths};
  \node[boxy, left=0pt of N] (NewAbs)
    {New abstraction\\Number as a\\measurement relation};

  \coordinate (SR) at (4.2,0);
  \coordinate (NL) at (-4.2,0);

  \draw[arrow] (Abstract) -- (SR);
  \draw[arrow] (SR) -- (Concrete);
  \draw[arrow] (Concrete) -- (NL);
  \draw[arrow] (NL) -- (Abstract);
\end{tikzpicture}
}
\caption{Example of the dialectical learning cycle in Davydov’s and Gattegno's approach to the concept of number.}
\label{number-cycle}
\end{figure}

V. V. Davydov’s educational project, rooted in dialectical materialism, conceives of the
relationship between the abstract and the concrete as a dynamic and reciprocal movement.
Contrary to traditional accounts that begin with concrete, empirical objects and ascend toward
abstract generalities, Davydov proposes a path that \emph{begins from the abstract}, understood
as the essential relations constitutive of a concept. For instance, in the case of number, he
starts not with counting objects but with relations of order and equivalence. These are mediated
through symbolic forms, enacted in concrete activities, and ultimately yield a new, more
refined abstraction: number as a relation of measurement.  \cite{Coles:106mf} notes the similarity of the Davydov influenced  
US Measure Up programme and Gattegno Mathematics. This non-traditional progression is illustrated in Figure~\ref{number-cycle}.

George \cite{lakoff1987women} introduced the study of the mind within the emerging field of cognitive science: integrating psychology, linguistics, anthropology, philosophy, and computer science. The core questions involve the nature of reason, conceptual systems, catagorisation, and whether thought is universal or shaped by human embodiment. Lakoff contrasts two major perspectives. The traditional view, or \emph{objectivism}, sees reason as abstract, disembodied, and literal. It treats thought as the mechanical manipulation of symbols, modelled on formal logic and computation, where meaning arises from a direct correspondence between symbols and an objective world. Categories, in this framework, are defined by shared properties and necessary and sufficient conditions.  

The new perspective, or \emph{experiential realism}, proposes that reason is embodied, imaginative, and rooted in human perception, bodily experience, and social interaction. Categorisation is not purely logical but relies on prototypes, basic-level categories, metaphor, metonymy, and imagery. Human reason is not a version of transcendental reason but instead develops from lived experience.  Lakoff stresses that this shift has broad implications. If reason is understood as disembodied and mechanical, human intelligence, creativity, and art are devalued, and the mind is reduced to the status of a computer. This is a real and present risk when influential software vendors describe statistical prediction as \textit{thinking}. Recognising reason as embodied and imaginative instead enriches our understanding of learning, creativity, and humane living.

The category theorist Charles \cite{Wells:kx} argues that mathematics instruction and exposition can be enriched by adopting habits and metaphors from computer science. He argues that mathematicians would do well to borrow the practice of specifying external behaviour from computer science. For example, a program for factoring large integers might use a
function \texttt{PrimeQ :: $\mathbb{Z}$ -> Bool} with the property that \texttt{PrimeQ n}
returns \texttt{True} if the integer \texttt{n} is prime and \texttt{False} otherwise. This description gives
the function's external behaviour. It says nothing about how that behaviour is
implemented.

The computer science notion of types and polymorphism also sheds light on
mathematics education. Just as a compiler flags a type error, students should be
taught to recognise and avoid type confusions---for example, conflating a
function with its defining expression or its values. Explicit attention to types
clarifies the multiple meanings of symbols such as $\times$, which varies across
contexts from integer multiplication to composition of fractions as operators to cartesian product.

Wells draws on hacker culture to suggest ways mathematicians might better communicate and reflect on their own practices.\footnote{A hacker is someone who programs for the pleasure of programming---although
useful software may result, this is not their primary motivation. A hacker enjoys
learning and using the esoteric features and behaviour of operating systems and
programming languages. The later media usage as someone who breaks into private
systems is \emph{not} intended here.}
He describes the state of keeping many details in mind while programming as \textit{juggling eggs}. He notes that this has a clear analogue in mathematics. It is a phenomenon familiar to research mathematicians. You can’t
spend short, separated pieces of time trying to understand a complicated mathematical
phenomenon; you need the time to concentrate and to get it all in your head at once
to be in \textit {``hacking''} and \textit {``debugging mode.''}

\cite{Rule:2020uk}  illustrate the hacker metaphor through children's strategies in small-number addition.
Empirical studies show that children do not simply learn one canonical addition strategy; instead
they experiment with multiple techniques (counting on fingers, decomposition, retrieval from
memory, exploiting commutativity). For example, in solving $5+3$, a child might recall $3+5$
instead, recognising equivalence and saving effort. Such `hacks' demonstrate opportunistic use
of available knowledge, not just systematic hypothesis testing.

Computer science and mathematics are mutually reinforcing in K–12 education. A systemic review by \cite{schulte2024review} shows growing attention to their integration via computational thinking, algorithms, modelling, and data sciences. They conclude that realising this promise at scale will require coherent definitions, robust
assessments, inclusive pedagogies, and sustained teacher learning. With careful design, students
can experience mathematics not merely as static symbol manipulation but as a dynamic,
computational medium for thinking and creating.

\subsection{Mathematics Circles}
The system of mathematical circles in Russia has become a remarkable mass precedent of the Learning-by-doing principle. 

It was instigated in the circle of Prof. Nikolai Luzin at Moscow University in 1920-30. In this circle, young researchers were directly involved in the creation of new mathematics, which was the main part of their mathematical education. 

In the mid-1930s, this tradition was transferred by leading Russian mathematicians to the university level in circles and to Olympiads for schoolchildren in Moscow, Leningrad and other cities. It was continued after the war. 

Since the 1960s there have been two important extensions of the circle tradition. The first was that mathematical schools were established, including boarding schools, associated with major universities. In these schools, mathematics classes in the Learning-by-doing model became part of the core curriculum, where it was also supplemented by programming. The key figures here were Andrei Kolmogorov, Alexander Kronrod and Nikolai Konstantinov \citep{Konstantinov:2021aa}. The second was the creation of distance learning circles, a correspondence mathematical school, where a world–renowned mathematician, Israel Gelfand, also played a key role. 

Since the 1990s the circle system has become popular in the United States, largely with the participation of Russian immigrants. In recent years, thanks to the initiative of Alexander Gerko, a graduate of the system of mathematical circles and schools, this system has been working effectively in British education.

\section{Cuisenaire-Gattegno: An action research tradition in mathematics learning}
\label{action}

Caleb Gattegno worked with pure mathematicians Jean Dieudonn\'e, Gustave Choquet and colleagues in the Bourbaki group to apply mathematical thinking through action research to the design of educational processes \citep{Benson:2022xi}. Action research is a classroom based, iterative process where researchers and practitioners work together to identify issues, gather data, implement solutions,  evaluate and peer-review results. 

Choquet placed close attention to experiments in teaching young children an enriched technical vocabulary with Cuisenaire rods -- the manipulative technology invented by Georges Cuisenaire and further developed by Gattegno. Cuisenaire's rods are cuboids, each the length of a multiple of the length of the smallest - a 1cm white cube. Rods of the same size have the same colour and vice versa. Choquet ``became both adept at the approach and a skilled user of the rods'' \citep{Choquet:1963ve}. 

On the basis of this experience Gattegno claimed that he could cover the `counting led' curriculum in eighteen months.

\subsection{Equivalence and Transformation}

From the outset Gattegno introduces the concept of \textit{equivalence} as a generalisation of \textit{equivalent colour} and \textit{equivalent length}. He remarked that each of us learnt to talk before or around the age of two and equivalence played a key part of this learning. ``The mental equipment for mastering (speech) must actually exist and it proves its existence by functioning. \dots One of the difficulties (in modelling this mathematically) resides in the fact that the grasp of meanings precedes verbalisation and that words per se are not the message, but only one of the possible vehicles for the message. \dots To have a working model---a mathematics---it is necessary to \dots reach the way in which meanings select their own expressions and place them adequately in the flow of speech in order to provide, through a set of \textit{transformations}, the required equivalences. It is my hunch today that \textit{equivalence}, which carries within itself the dynamic component of transformation, is the cardinal concept of mathematics. Perhaps one day I shall be able to tell the story of `equivalence through transformation’ and its place in the study of speech.''
\citep[p.~136]{Gattegno:1970ow}.

%
%
%

Gattegno uses operations with the rods -- placing them end to end, side by side or stacked as towers -- to model sets with structure such as the integer and rational number systems.  In this approach ``all the operations with integers and fractions can be studied simultaneously (with coloured rods); whole numbers being recognised as the equivalence class of their partitions and fractions as ordered pairs, one serving to measure the other, or as operators belonging to classes of equivalence which are the rational numbers involved in the operations'' \cite[p. 201]{Fedon:1966ve}. He demonstrated that ``Children of six or seven are thoroughly familiar with their tables, children of five conceive and compare fractions easily and accurately, children of eight solve simultaneous equations and at 10 they understand permutations and combinations which they themselves form and analyse'' \citep[p. 88]{Gattegno:1956jb}.

\section{Computational Conceptual Mathematics}
\label{haskell}

In the early 1950s Gattegno and his colleagues set up to a Commission to reform school mathematics  \citep{C.-Gattegno:1965hc,C.-Gattegno:1965zr,Benson:2010vn}. Choquet's \textit{What is Modern Mathematics} became a manifesto for their work. Chapter 1 was entitled the ``Axiomatic method: Structures, Characteristics and Dangers.'' Chapter 2 covered some the tools of conceptual mathematics: \textit{morphisms,  categories and functors}. 
Introducing \cite{Choquet:1963ve} Gattegno wrote, ``the pithy presentation which follows will help readers ...  to think more adequately of the articulation of the research of today into the teaching of today and tomorrow. Its value for readers cannot but increase as they reach deeper and deeper understanding of the various sections.'' Choquet wrote ``Since Bourbaki has such clear-cut concepts and is so intimately associated with the development of mathematics in our time, we can hope that a study of  `his' philosophical and mathematical work may lead us to the essence of modern trends in analysis. Such a study may serve to develop for all levels of education a teaching of mathematics better adapted to the needs of our time and the level of awareness of our generation.'


The traditional approach to school mathematics is to start with counting and arithmetic and proceed through three distinct aspects: algebra – the manipulation of symbols, geometry – dealing with shape and position, and logic – making arguments. Reenforced with drill and practice this curriculum has remained largely unchanged since the nineteenth century. As we have noted, it underserves a third of the school population in England. Many will  fall behind expected standards at the end of primary school. Most fail to recover this lost ground in secondary school.

Gattegno's non-traditional approach, founded on conceptual mathematics, or category theory, combines algebra with geometry and logic. It is about the structure of arguments, and deals with algebra geometrically. While category theory only succeeds in making small portions of mathematics easy — these are the portions that lie closest to the core of the subject, the part that illuminates the rest. It lies at the heart of a revolution in how pure mathematicians do research. Since the turn of the century category theory has developed direct applications in ecological diversity, engineering, biology, chemistry, theoretical physics, computer science and informatics. In this paper we address the challenges of applying it to overcome the limitations of the traditional \textit{instruct and perform} model of mathematics education. Our proposal is to marry it with computational tools, such as generative AI agents and software supported type inference in the functional programming language Haskell. This approach shows promise in overcoming the teacher education barrier, one of several factors that holds back Gattegno's initiative from wider adoption.

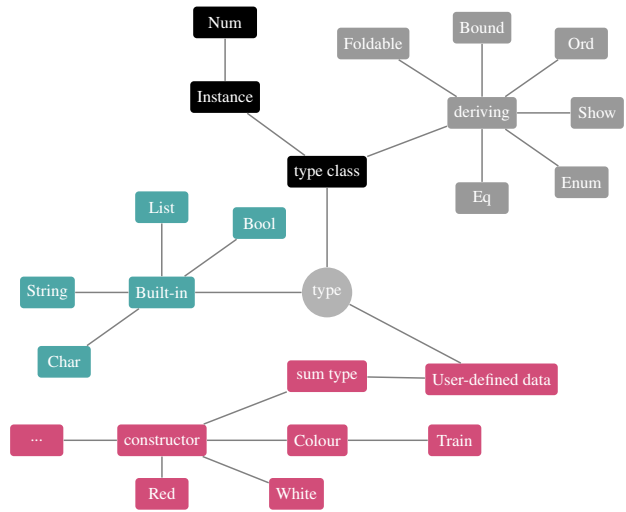
\begin{figure}[htbp]
\centering
\scalebox{0.7}{
\begin{tikzpicture}[
    node distance=1.5cm,
    every node/.style={text=white, font=\footnotesize},
    builtin/.style={fill=teal!70, rectangle, rounded corners=2pt, minimum width=1cm, minimum height=0.6cm},
    typeclass/.style={fill=black, rectangle, rounded corners=2pt, minimum width=1.2cm, minimum height=0.6cm},
    deriving/.style={fill=gray!80, rectangle, rounded corners=2pt, minimum width=1cm, minimum height=0.6cm},
    userdef/.style={fill=purple!70, rectangle, rounded corners=2pt, minimum width=1.5cm, minimum height=0.6cm},
    constructor/.style={fill=purple!70, rectangle, rounded corners=2pt, minimum width=1cm, minimum height=0.6cm},
    center/.style={fill=gray!60, circle, minimum size=0.8cm},
    arrow/.style={-, thick, gray}
]

\node[center] (type) {type};

\node[builtin, left=2cm of type] (builtin) {Built-in};
\node[builtin, above=1cm of builtin] (list) {List};
\node[builtin, above right=0.7cm and 0.7cm of builtin] (bool) {Bool};
\node[builtin, left=1cm of builtin] (string) {String};
\node[builtin, below left=0.7cm and 0.7cm of builtin] (char) {Char};

\node[typeclass, above=1.5cm of type] (typeclass) {type class};
\node[typeclass, above left=0.8cm and 0.5cm of typeclass] (instance) {Instance};
\node[typeclass, above=0.8cm of instance] (num) {Num};

\node[deriving, above right=0.5cm and 1.5cm of typeclass] (deriving) {deriving};
\node[deriving, above=1cm of deriving] (bound) {Bound};
\node[deriving, above right=0.7cm and 0.7cm of deriving] (ord) {Ord};
\node[deriving, right=1cm of deriving] (show) {Show};
\node[deriving, below right=0.7cm and 0.7cm of deriving] (enum) {Enum};
\node[deriving, below=1cm of deriving] (eq) {Eq};
\node[deriving, above left=0.7cm and 0.7cm of deriving] (foldable) {Foldable};

\node[userdef, below right=1cm and 1.5cm of type] (userdef) {User-defined data};
\node[userdef, below=0.8cm of type] (sumtype) {sum type};
\node[constructor, below left=0.6cm and 1.5cm of sumtype] (constructor) {constructor};
\node[constructor, left=1cm of constructor] (dots) {...};
\node[constructor, below=0.4cm of constructor] (red) {Red};
\node[constructor, right=1.5cm of red] (white) {White};
\node[constructor, right=1.5cm of constructor] (colour) {Colour};
\node[constructor, right=1.5cm of colour] (train) {Train};

\draw[arrow] (builtin) -- (type);
\draw[arrow] (list) -- (builtin);
\draw[arrow] (bool) -- (builtin);
\draw[arrow] (string) -- (builtin);
\draw[arrow] (char) -- (builtin);

\draw[arrow] (typeclass) -- (type);
\draw[arrow] (instance) -- (typeclass);
\draw[arrow] (num) -- (instance);

\draw[arrow] (deriving) -- (typeclass);
\draw[arrow] (bound) -- (deriving);
\draw[arrow] (ord) -- (deriving);
\draw[arrow] (show) -- (deriving);
\draw[arrow] (enum) -- (deriving);
\draw[arrow] (eq) -- (deriving);
\draw[arrow] (foldable) -- (deriving);

\draw[arrow] (userdef) -- (type);
\draw[arrow] (sumtype) -- (userdef);
\draw[arrow] (constructor) -- (sumtype);
\draw[arrow] (dots) -- (constructor);
\draw[arrow] (red) -- (constructor);
\draw[arrow] (white) -- (constructor);
\draw[arrow] (colour) -- (constructor);
\draw[arrow] (colour) -- (train);

\end{tikzpicture}
}
\caption{Haskell Type System Hierarchy for Colour and Train }
\label{fig:haskell-concept-graph}
\end{figure}

Today,  programming languages like Scratch and Python offer learners and teachers in primary school the opportunity to become acquainted with data types: characters (\texttt{`c'}), strings (\texttt{"hello world"}), booleans (\texttt{True/False}) and functions (\texttt{f x = x + 1}).  Type classes in Haskell abstract common functions found in built-in types such as \texttt{Char} and \texttt{String} and make this functionality available to new types. A user-defined data type in Haskell can inherit various attributes from a type class as illustrated in  Figure~\ref{fig:haskell-concept-graph}. For example, working with Cuisenaire rods is modelled with the \textit{sum data type} \texttt{Colour}. This is a set of distinct \textit{constructors}: \texttt{White, Red} etc, one for each colour. By instantiating (deriving) type classes such as \texttt{Eq} and \texttt{Ord}, elements of type \texttt{Colour} can be compared (for equality) and ordered (\texttt{Ord})  \citep{Benson:2017pt}.

Our classroom research projects and text-books, discussed below,  build on this familiarity to offer teachers an enriched technical vocabulary: including \textit{sorting, naming, set, equivalence, permutation, combination, partition, number system, bag, data type, group, vector, matrix} and \textit{linear algebra}. We use Haskell as a lingua franca  to give an executable, computational, dimension to their studies \citep{Hudak:2000aa}. Where traditional programming forces the learner to work with a model of computation close to that of a Von Neumann machine, Haskell's model is nearer to conceptual mathematics proper. \textbf{Cat Hask} is the informal name for the category of Haskell data types and functions \citep{Fong:aa}.

\section{Related Work}
\label{others}

\subsection{Mathematics and Informatics}

In recent decades, several Russian schools have been implementing a world-unique education program of mathematics for elementary schools. In it, the landscape of school arithmetic is radically expanded
due to the basic objects of modern mathematics and computer science. \textit{Virtual actions} on visual objects make these concepts much easier to understand. The range of activities also
expands due to, for example, the introduction of strategies for enumeration, game winning, and algorithms (also operating in a visual environment). At the same time, the student’s position changes: they independently discover and build mathematics and constantly solve personally new, but feasible tasks that often do not look familiar. Exercises are contained in a series of textbooks for Grades 1 -- 4 of primary school. Structured drawing represents strings, bags, tables, cycles, trees. By working with these objects students develop their ability to reason mathematically \citep{Posicelskayaa:2023aa}.

%
%
%
%
%
%

\subsection{From visual objects to dynamic programming}

Structured drawing of visual objects can provide a gentle, visual path from functional into imperative and object-oriented programming and program design. Karel the Robot illustrates one such pathway. Karel is an educational programming environment devised by Richard E. Pattis in the early 1980s at Stanford to teach core computing ideas without the overhead of a full programming language. Named after writer Karel Čapek (who popularised the word “robot”), Karel lives on a 2-D grid of “streets” and “avenues,” interacting with walls and simple objects called beepers. Students write tiny programs using a small, readable vocabulary—commands like move, turnLeft, putBeeper, and pickBeeper—combined with control structures (loops, conditionals, and defining new functions) and `sensory' tests such as frontIsClear. Joe Bergin has developed Monty Karel as an introduction to computer programming for novices. It uses the Python programming language to introduce the principles of dynamic object-oriented programming. A student able to do the exercises in this book, or one of its companions, well on his or her way to a deep understanding of programming  \citep{Bergin:2013aa}.

In parallel, in the late 1970s, the group of V. Betelin and A. Kushnirenko developed a similar software environment for teaching computer university students. In the very first versions, commands for the robot were set using separate punched cards, and the result of the program execution – the trajectory of the robot - was printed on an alphanumeric printer  \citep{Betelin:2020aa}. In 1985, such a computer environment was used to teach programming to all high school students in the Soviet Union on personal computers, and in the late 1980s it was moved to elementary schools and kindergartens \citep{Kushnirenko:2023aa}).

\section{Co-designing a new learning model}
\label{codesign}

Gattegno argued that the primary role of software is to transform mental presence \citep{Gattegno:1981kz,Benson:2026aa}. The computer trains alertness. In this section we explore the implications of these findings for educating learner awareness of their internal dialogue, and teacher participation in the co-creation of software supported educational mathematics.

Mathematics is a systematic domain of human activity, where the objects are formal structures and the rules for manipulating them. The challenge posed by educational mathematics software is not simply to accurately reflect existing domains, but to allow designers to create new ones. Winograd and Flores, in \textit{Understanding Computers and Cognition: A New Foundation for Design} set out an approach to software design that emphasised looking for recurrent patterns of conversation, rather than regularity in data alone. They argue that patterns exist in the domain of conversation, not in some mental domain of the participants. They wrote, ``when we are engaged in successful language activity, the conversation is not present-at-hand, as something observed. We are immersed in its unfolding. Its structure becomes visible only when there is some kind of breakdown'' \citep{Winograd:1986lm}.

In our approach teachers and teacher-interlocutors, equipped with appropriate technical vocabulary and notations, learn how to analyse and construct mathematising situations both face-to-face and at a distance. They ask what are the types of data in a problem? what are the structural relationships between the data? what are the properties of the data and the structures? And they ask what are the affordances of educational mathematics software? how can we reorganise the processes of mathematics teaching and learning to take advantage of them? how can teachers learn alongside their students? how can they quality assure their work?

The learners need, in John Mason's words,  to develop an ``inner witness or monitor which observes without involvement while the predicated `I' is immersed, fully involved and caught up in action.  This is much more than post-event introspection or reflection. Its origins are ancient. For example, it is hinted at by the image of two birds found in the ancient text the Rg Veda (Samhita 1.164.20)''  \citep{Mason:2004ty}:

\begin{verbatim}
“Two birds, close-yoked companions, 
both clasp the self-same tree;
one eats of the sweet fruit,
the other looks on without eating.”
\end{verbatim}

\subsection{Teacher/Researcher as co-designer: Researcher/Interlocutor as consultant }

It is in this spirit that one author has been researching the new division of labour in the network enabled workplace \citep{Ian-Benson:1983qf,Benson:1986px,Benson:2014ys}. In such an extended classroom students are programmers,  teachers learn alongside them, and the reseacher/interlocator engages directly:  giving feedback to the learners, and co-designing lessons and lesson sequences with the teacher. Type inference with an interactive programming environment helps to educate the learner's awareness of infinitesimal steps in reasoning. Giving learners the tools to validate and verify their reasoning is a necessary complement to statistically predicted code.

%
%

\section{Conclusion: Agents and Agency}
\label{conclude}

Writing fifty years ago Gattegno remarked on teacher agency, `` Today we have the means to criticise tests seriously and to replace them with activities that truly reflect what learners do with themselves and also give teachers meaningful information on how to steer courses and develop realistic curricula. The need for knowing what one is doing exists all the time and only a cybernetic approach to the process of learning with continuous control via conscious criteria can be satisfactory to teachers, students, and public alike. The role of the teacher will then be elevated to that of a scientist, thus permitting teachers to form a responsible profession and to deliver to each generation what it needs in order to meet its future'' \citep{Gattegno:1970tg}.  Steve \cite{Watson:2025aa}  have argued that  ``the integration of generative AI into scientific authorship requires a continuous negotiation of roles and boundaries, leading to a symbiotic relationship where both AI and human authors, but also social systems contribute to the final scientific output. This co-evolution reflects a dynamic where both entities influence and shape each other, with scientists adapting their practices to incorporate AI and AI evolving through its interactions with scientists.''

We have shown how polymorphic type inference, as anticipated by Charles Wells, can help learners to become self-aware. The same technical infrastructure can enable teachers to participate as researchers in the redesign of mathematics learning networks.  John Mason and Anne Watson have studied the role of AI agents in mathematics education. They write, ``the onus on teachers, and the aims of education generally, are likely to be more and more centred on the need, and the means, to interrogate expertise, whether manifested as a chatbot or as an actual person. As AI takes over more and more of the routine functioning of human beings, and as long as the exorbitant costs of energy and cooling are tolerated, education will turn away from its current addiction to performance, and towards probing, interrogating, and doubting \citep{Mason:2026aa}. 

For this enterprise to be successful the nature of the negotiation between AI agents, teachers and teacher/interlocutors needs to become explicit as suggested by Winograd and Flores. While AI agents have a role to play in educating mathematics teachers' awareness of their own mental powers this will only happen if we cease to anthropomorphesise  software \citep{Benson:1971bh}. If the results of these agents' textual predictions are to be of use to teachers they must be embedded deep down in a network of negotiated interactions: their outputs subordinate to verification with Interactive Development Environments such as iHaskell or computer algebra systems such as Sage Math, Mathematica or the Lean proof assistant.

Teachers and teacher/interlocutors as researchers have a role to play is seeding and nurturing such networks. \cite{Winograd:1986lm} argue that we are primarily engaged ``in a philosophical discourse about the self—about what we can do and what we can be. Tools are fundamental to action, and through our actions we generate the world. The transformation we are concerned with is not a technical one, but a continuing evolution of how we understand our surroundings and ourselves -- of how we continue becoming the beings that we are.''

The transition to a new way of working should be co-designed with teachers in a gradual passage to a new model. This will be characterised by an attitude of a teacher as a lifelong learner, alongside their learners and from them. In mathematics this process can go in parallel with the inclusion of future teachers in the work of mathematics circles and pilot schools, where such a model is already being implemented.

\bibliographystyle{plainnat}
\bibliography{arxiv}

\begin{thebibliography}{48}
\providecommand{\natexlab}[1]{#1}
\providecommand{\url}[1]{\texttt{#1}}
\expandafter\ifx\csname urlstyle\endcsname\relax
  \providecommand{\doi}[1]{doi: #1}\else
  \providecommand{\doi}{doi: \begingroup \urlstyle{rm}\Url}\fi

\bibitem[ATM(2023)]{Group:2023aa}
ATM.
\newblock {R}esponse to {R}oyal {S}ociety {M}athematics {F}utures
  {C}onsultation, 2023.
\newblock URL
  \url{https://atm.org.uk/write/MediaUploads/Consultations/ATMCTMMFResponse.pdf}.

\bibitem[Bakhtin(1981)]{BakhtinDialogic}
Mikhail~M. Bakhtin.
\newblock \emph{The Dialogic Imagination: Four Essays}.
\newblock University of Texas Press, 1981.

\bibitem[Bakhtin(1984)]{BakhtinPoetics}
Mikhail~M. Bakhtin.
\newblock \emph{Problems of Dostoevsky's Poetics}.
\newblock University of Minnesota Press, 1984.

\bibitem[Benson(1971)]{Benson:1971bh}
Ian Benson.
\newblock Machines that mimic thought.
\newblock \emph{New Scientist}, 51\penalty0 (767), 2 Sept 1971.

\bibitem[Benson(1986)]{Benson:1986px}
Ian Benson, editor.
\newblock \emph{Intelligent Machinery: Theory and Practice}.
\newblock Cambridge University Press, 1986.

\bibitem[Benson(2010)]{Benson:2010vn}
Ian Benson.
\newblock Letter from {W}hitehall.
\newblock https://www.cl.cam.ac.uk/downloads/ring/ring-2010-01.pdf, January
  2010.

\bibitem[Benson(2014)]{Benson:2014ys}
Ian Benson.
\newblock Can computer science rescue mathematics reform?
\newblock http://www.cl.cam.ac.uk/downloads/ring/ring-2014-09.pdf, 2014.

\bibitem[Benson and Cane(2017)]{Benson:2017pt}
Ian Benson and Jenny Cane.
\newblock \emph{Hello {W}orld}, volume~2, chapter Using {H}askell with 5- to 7-
  year olds, pages 60--61.
\newblock Computing at {S}chool, Summer 2017.

\bibitem[Benson and Lloyd(1983)]{Ian-Benson:1983qf}
Ian Benson and John Lloyd.
\newblock \emph{New {T}echnology and {I}ndustrial {C}hange: The {I}mpact of the
  {S}cientific-{T}echnical {R}evolution on {L}abour and {I}ndustry}.
\newblock Kogan Page (London), Nichols (New York), 1983.

\bibitem[Benson and Singer(2026)]{Benson:2026aa}
Ian Benson and Jeremy Singer.
\newblock Haskell in {S}chool: {F}unctional {P}rogramming for {S}chool {A}ge
  {L}earners.
\newblock \emph{(under submission)}, 2026.

\bibitem[Benson et~al.(2022)Benson, Marriott, and McCandliss]{Benson:2022xi}
Ian Benson, Nigel Marriott, and Bruce~D. McCandliss.
\newblock Equational reasoning: A {S}ystematic {R}eview of the
  {C}uisenaire-{G}attegno {A}pproach.
\newblock \emph{Frontiers in Education
  https://doi.org/10.3389/feduc.2022.902899}, 7:902899, 2022.
\newblock URL \url{https://doi.org/10.3389/feduc.2022.902899}.

\bibitem[Bergin et~al.(2013)Bergin, Stehlik, Rogers, and Pattis]{Bergin:2013aa}
J.~Bergin, M.~Stehlik, J.~Rogers, and R.~Pattis.
\newblock \emph{Monty {K}arel:{A} {G}entle {I}ntroduction to the {A}rt of
  {D}ynamic {O}bject-{O}riented {P}rogramming in {P}ython}.
\newblock 2013.

\bibitem[Betelin et~al.(2020)Betelin, Kushnirenko, and Leonov]{Betelin:2020aa}
V.~B. Betelin, A.~G. Kushnirenko, and A.~G. Leonov.
\newblock Basic concepts of programming expounded for preschoolers.
\newblock \emph{Informatics and Applications [in Russian]}, 14\penalty0
  (3):\penalty0 55--61., 2020.
\newblock URL \url{https://doi.org/10.14357/19922264200308}.

\bibitem[Bibler(2009)]{Bibler:2009aa}
Vladimir~Solomonovich Bibler.
\newblock The foundations of the school of the dialogue of cultures program.
\newblock \emph{Journal of Russian and East European Psychology}, 47\penalty0
  (1):\penalty0 34--60, 2009.

\bibitem[Choquet(1963)]{Choquet:1963ve}
Gustave Choquet.
\newblock \emph{What is Modern Mathematics?}
\newblock Educational Explorers, 1963.

\bibitem[Coles(2021)]{Coles:106mf}
Alf Coles.
\newblock Commentary on a {S}pecial {I}ssue: {D}avydov's approach in the
  {X}{X}{I} century: {V}iews from {M}ultiple {P}erspectives.
\newblock \emph{Educational Studies in Mathematics}, 106\penalty0 (3):\penalty0
  471--478., 2021.

\bibitem[Comenius(1907)]{Comenius1907}
John~Amos Comenius.
\newblock \emph{The Great Didactic of John Amos Comenius}.
\newblock A. \& C. Black, 1907.
\newblock Originally written in the 17th century.

\bibitem[Fedon(1966)]{Fedon:1966ve}
J.~P. Fedon.
\newblock A study of the {C}uisenaire-{G}attegno method as opposed to an
  eclectic approach for promoting growth in operational technique and concept
  maturity with first grade children.
\newblock Master's thesis, Temple University, 1966.

\bibitem[Fong et~al.(2020)Fong, Milewski, and Spivak]{Fong:aa}
Brendan Fong, Bartosz Milewski, and David~I. Spivak.
\newblock Programming with categories (draft), 2020.
\newblock URL
  \url{http://brendanfong.com/programmingcats_files/cats4progs-DRAFT.pdf}.

\bibitem[Gattegno and et~al(1965)]{C.-Gattegno:1965zr}
C.~Gattegno and W.~Servais et~al.
\newblock Tome 2 etude de materiel.
\newblock In \emph{L`enseignement des mathematics}. Delachaux et Niestle, 1965.

\bibitem[Gattegno et~al.(1965)Gattegno, Piaget, Beth, Dieudonn\'e,
  Lichnerowicz, and Choquet]{C.-Gattegno:1965hc}
C.~Gattegno, J.~Piaget, E.W. Beth, J.~Dieudonn\'e, A.~Lichnerowicz, and
  G.~Choquet.
\newblock \emph{L enseignement des mathematics: Tome 1 Nouvelle Perspectives}.
\newblock Delachaux et Niestle, 1965.

\bibitem[Gattegno(1956)]{Gattegno:1956jb}
Caleb Gattegno.
\newblock {N}ew {D}evelopments in {A}rithmetic {T}eaching in {B}ritain:
  {I}ntroducing the {C}oncept of `{S}et'.
\newblock \emph{Arithmetic Teacher}, 3\penalty0 (3):\penalty0 85--89, April
  1956.

\bibitem[Gattegno(1963)]{Gattegno:1963cr}
Caleb Gattegno.
\newblock \emph{Mathematics with Numbers in Colour: Numbers from 1 to 20},
  volume~I.
\newblock Educational Explorers, Fishguard, 1963.

\bibitem[Gattegno(1970{\natexlab{a}})]{Gattegno:1970ow}
Caleb Gattegno.
\newblock The human element in mathematics.
\newblock In {ATM members}, editor, \emph{Mathematical Reflections}, pages
  131--137. CUP, 1970{\natexlab{a}}.

\bibitem[Gattegno(1970{\natexlab{b}})]{Gattegno:1970tg}
Caleb Gattegno.
\newblock \emph{What we owe children: the subordination of teaching to
  learning}.
\newblock Outerbridge and Dienstfrey, New York, 1970{\natexlab{b}}.

\bibitem[Gattegno(1981, 2009)]{Gattegno:1981kz}
Caleb Gattegno.
\newblock \emph{Computer and the Mind}, volume XI:1.
\newblock Newsletter: Educational Solutions, September 1981, 2009.

\bibitem[Gibb and Peal(2025)]{Gibb:2025aa}
Nick Gibb and Robert Peal.
\newblock \emph{Reforming {L}essons: {W}hy {E}nglish {S}chools {H}ave
  {I}mproved {S}ince 2010 and How This Was Achieved}.
\newblock Routledge, 2025.

\bibitem[Hadamard(1945)]{Hadamard1945}
Jacques Hadamard.
\newblock \emph{The Psychology of Invention in the Mathematical Field}.
\newblock Princeton University Press, 1945.

\bibitem[Halmos(1975)]{HalmosMoore}
Paul~R. Halmos.
\newblock The problem of learning to teach.
\newblock \emph{American Mathematical Monthly}, 82\penalty0 (5):\penalty0
  466--470, 1975.

\bibitem[Hudak et~al.(2000)Hudak, Peterson, and Fasel]{Hudak:2000aa}
Paul Hudak, John Peterson, and Joseph Fasel.
\newblock A {G}entle {I}ntroduction to {H}askell98.
\newblock 2000.

\bibitem[Keating(2026)]{Keating:2026aa}
Rose Keating.
\newblock Key {S}tage 2 {M}athematics {T}est {I}tem {P}erformance {A}nalysis.
\newblock \emph{Mathematical Teaching}, 2026.

\bibitem[Konstantinov and Semenov(2021)]{Konstantinov:2021aa}
N.~N. Konstantinov and A.~L. Semenov.
\newblock Productive education in mathematical schools.
\newblock \emph{Doklady Mathematics}, 106:\penalty0 S270--S287, 2021.

\bibitem[Kushnirenko et~al.(2023)Kushnirenko, Leonov, and
  Polikarpov]{Kushnirenko:2023aa}
A.~G. Kushnirenko, A.~G. Leonov, and S.~A. Polikarpov.
\newblock Error-{F}ree 2d {P}ictogrammic {S}yntax in a {P}rogramming {L}earning
  {E}nvironment for preschool children.
\newblock In Radek Silhavy, Petr Silhavy, and Zdenka Prokopova, editors,
  \emph{Software Engineering Application in Systems Design}, pages 637--651,
  Cham, 2023. Springer International Publishing.
\newblock ISBN 978-3-031-21435-6.

\bibitem[Lakoff(1987)]{lakoff1987women}
George Lakoff.
\newblock \emph{Women, Fire, and Dangerous Things: What Categories Reveal about
  the Mind}.
\newblock The University of Chicago Press, Chicago, 1987.
\newblock ISBN 0-226-46804-6.

\bibitem[Land and Benson(2024)]{Land:2024aa}
Frank Land and Ian Benson.
\newblock Letter: Tory schools improvement is far from the whole story.
\newblock \emph{FT}, March 11 2024.

\bibitem[Marks et~al.(2023)Marks, Barclay, and Barnes]{Marks:2023aa}
Rachel Marks, Nancy Barclay, and Alison Barnes.
\newblock \emph{The {P}revalence and {U}se of {T}extbooks and {C}urriculum
  {R}esources in {P}rimary {M}aths}.
\newblock Nuffield Foundation, 2023.

\bibitem[Mason(2004)]{Mason:2004ty}
John Mason.
\newblock Doing \( \neq \) construing, and doing + discussing \( \neq \)
  learning: The importance of the structure of attention.
\newblock In \emph{ICME10}, 2004.

\bibitem[Parker and Mahavier(1992)]{MahavierMore}
Gary~E. Parker and William~S. Mahavier.
\newblock Getting {M}ore from {M}oore.
\newblock \emph{Primus}, 2:\penalty0 235--246., September 1992.
\newblock Online educational essay.

\bibitem[P{\'o}lya(1945)]{Polya1945}
George P{\'o}lya.
\newblock \emph{How to Solve It: A New Aspect of Mathematical Method}.
\newblock Princeton University Press, 2004 edition, 1945.

\bibitem[Posicelskayaa et~al.(2023)Posicelskayaa, Rudchenkob, and
  Semenov]{Posicelskayaa:2023aa}
M.~A. Posicelskayaa, T.~A. Rudchenkob, and A.~L. Semenov.
\newblock Mathematical elements of elementary education.
\newblock \emph{Doklady Mathematics}, Vol. 107\penalty0 (Suppl. 1):\penalty0
  S10--S41, 2023.
\newblock URL \url{https://bit.ly/MathElements}.

\bibitem[Rule et~al.(2020)Rule, Tenenbaum, and Piantadosi]{Rule:2020uk}
Joshua~S. Rule, Joshua~B. Tenenbaum, and Steven~T. Piantadosi.
\newblock The {C}hild as {H}acker.
\newblock \emph{Trends in Cognitive Sciences}, 24\penalty0 (11):\penalty0
  900--915, November 2020.
\newblock URL \url{https://doi.org/10.1016/j.tics.2020.07.005}.

\bibitem[Schulte and colleagues(2024)]{schulte2024review}
Carsten Schulte and colleagues.
\newblock What we talk about when we talk about k--12 computing education: A
  systematic review.
\newblock In \emph{ITiCSE 2024 Working Group Reports on Innovation and
  Technology in Computer Science Education}. Association for Computing
  Machinery, 2024.

\bibitem[Soprunova et~al.(2021-2021 In {R}ussian)Soprunova, Posicelskaya,
  Posicelsky, Rudchenko, and Semenov]{Soprunova:2021aa}
N.~A. Soprunova, M.A. Posicelskaya, S.E. Posicelsky, T.A. Rudchenko, and A.~L.
  Semenov.
\newblock \emph{Mathematics and {I}nformatics {G}rades 1 - 4: {T}extbook for
  {G}eneral {E}ducational {I}nstitutions}.
\newblock Moscow: Center for Pedagogical Excellence, 2021-2021 In {R}ussian.

\bibitem[Watson and Mason(2026)]{Mason:2026aa}
Anne Watson and John Mason.
\newblock Artificial {P}edagogic {I}ntelligence in the {C}ontext of
  {M}athematics {E}ducation.
\newblock \emph{forthcoming}, 2026.

\bibitem[Watson et~al.(2025)Watson, Brezovec, and Romic]{Watson:2025aa}
Steven Watson, Erik Brezovec, and Jonathan Romic.
\newblock The role of generative {A}{I} in academic and scientific authorship:
  an autopoietic perspective.
\newblock \emph{AI and Society}, 40:\penalty0 3225--3235, 2025.

\bibitem[Wegerif(2013)]{wegerif2013}
Rupert Wegerif.
\newblock \emph{Dialogic: Education for the Internet Age}.
\newblock Routledge, London, 2013.
\newblock ISBN 978-0-415-67407-4.

\bibitem[Wells(1995)]{Wells:kx}
Charles Wells.
\newblock Communicating mathematics:useful ideas from computer science.
\newblock \emph{The American Mathematical Monthly}, 102\penalty0 (5):\penalty0
  pp 397--408, May 1995.
\newblock URL \url{http://www.jstor.org/stable/2975030}.

\bibitem[Winograd and Flores(1986)]{Winograd:1986lm}
Terry Winograd and Fernando Flores.
\newblock \emph{{U}nderstanding {C}omputers and {C}ognition: {A} {N}ew
  {F}oundation for {D}esign}.
\newblock Ablex Pub. Corp., New Jersey, 1986.

\end{thebibliography}
\end{document}